# Extraordinary Photon Transport by Near-Field Coupling of a Nanostructured Metamaterial with a Graphene-Covered Plate

Jui-Yung Chang, Yue Yang, and Liping Wang*

School for Engineering of Matter, Transport & Energy
Arizona State University, Tempe, AZ 85287, USA
* Corresponding author: liping.wang@asu.edu

**Abstract**

Coupled surface plasmon/phonon polaritons and hyperbolic modes are known to enhance radiative transport across nanometer vacuum gaps but usually require identical materials. It becomes crucial to achieve strong near-field energy transfer between dissimilar materials for applications like near-field thermophotovoltaic and thermal rectification. In this work, we theoretically demonstrate extraordinary near-field radiative transport between a nanostructured metamaterial emitter and a graphene-covered planar receiver. Strong near-field coupling with two orders of magnitude enhancement in the spectral heat flux is achieved at the gap distance of 20 nm. By carefully selecting the graphene chemical potential and doping levels of silicon nanohole emitter and silicon plate receiver, the total near-field radiative heat flux can reach about 500 times higher than the far-field blackbody limit between 400 K and 300 K. The physical mechanisms are elucidated by the near-field surface plasmon coupling with fluctuational electrodynamics and dispersion relations. The effects of graphene chemical potential, emitter and receiver doping levels, and vacuum gap distance on the near-field coupling and radiative transfer are analyzed in detail.

# 1. INTRODUCTION

Near-field radiation has attracted much attention in the fields of energy harvesting [1, 2] and heat management [3-5] since it can exceed the far-field blackbody limit through coupling of evanescent waves [6, 7]. In particular, the near-field enhancement could be orders of magnitude with the excitation of surface plasmon/phonon polaritons (SPP/SPhP) across the nanometer vacuum gaps [8-11]; however, it usually requires identical or similar materials for the emitter and receiver to achieve the best coupling effect and thereby maximum heat flux enhancement. Recent studies on hyperbolic metamaterials (HMMs) [12-20], which exhibit large photonic local density of states, show the promise in enhancing the near-field radiative transfer by means other than resonant coupling, while the strong enhancement also requires matching hyperbolic behaviors for the emitter and receiver materials. It is still a challenge to greatly enhance the performance of near-field thermophotovoltaic (TPV) by either resonance coupling of SPP/SPhP or strong hyperbolic modes because of inherent mismatch in the dissimilar optical properties of the emitter and cell. TPV emitters are usually made of plasmonic metals or polar materials, while the cells are semiconductors with bandgap in the near infrared [21]. Therefore, it becomes crucial to find an efficient way to enhance near-field radiative transport between dissimilar materials.

Graphene, which supports surface plasmon [22, 23] with excellent tunability from near infrared to terahertz frequencies [24-26], has been recently studied and shown with capability to effectively modulate near-field radiative flux. By properly tuning the properties of graphene, near-field radiation between dielectrics can be significantly enhanced by covering graphene [27, 28]. Lim et al [29] theoretically showed enhanced or suppressed near-field radiative heat flux between two graphene-coated doped silicon plates, which varies with the silicon doping level and graphene chemical potential. Liu and Zhang [30] found more than one order of magnitude enhancement in near-field radiative flux between two corrugated silica gratings with graphene coating. By covering graphene on doped silicon nanowires, which exhibit strong hyperbolic behaviors, Liu et al [31] theoretically demonstrated near-unity photon tunneling probability in a broad frequency range and $k$ space by the coupling between graphene plasmon and hyperbolic modes. In terms of near-field radiative transport between dissimilar materials, Ilic et al [32] applied graphene on an emitter and showed that the near-field TPV system performance can be optimized by matching the graphene plasmon with the cell bandgap. Messina and Ben-Abdallah theoretically demonstrated improved near-field TPV efficiency between a hexagonal boron nitride (hBN) emitter and graphene-coated InSb cell by effective near-field coupling of hBN phonon modes with graphene plasmon [33].

In this study, we theoretically investigate the near-field radiative transport between a nanostructured metamaterial emitter made of doped silicon nanohole (D-SiNH) arrays and a doped silicon plate covered by monolayer graphene, as depicted in Fig. 1. The emitter and receiver, which are separated by a vacuum gap with distance $d$, are respectively maintained at $T_1$ = 400 K and $T_2$ = 300 K with doping levels $N_1$ and $N_2$. The SiNH emitter is described as an uniaxial homogeneous medium by effective medium theory (EMT) and graphene modifies the reflection coefficients at the vacuum-receiver interface as a surface current. Fluctuational electrodynamics incorporated with uniaxial wave propagation is employed to calculate the near-field radiative heat flux. The extraordinary enhancement in spectral near-field radiative transfer will be illustrated, while the underlying mechanism will be elucidated as unusual surface plasmon coupling between dissimilar materials with fluctuational electrodynamics and dispersion relation. The effects of graphene chemical potential, doping levels, and vacuum gaps on the near-field photon tunneling will be studied in detail as well.

## 2. THEORETICAL METHODS

### 2.1 Effective Dielectric Functions of Doped Silicon Nanohole Emitter

With the assumption that the feature size like hole array period $P$ is much smaller than the characteristic thermal wavelength, the D-SiNH emitter can be considered as a homogeneous uniaxial medium with effective dielectric functions described by the Maxwell-Garnett effective medium theory [34]:

$$\varepsilon_{\parallel,\text{eff}} = \frac{\varepsilon_{\text{Si}}(1+\varepsilon_{\text{Si}}) + f\varepsilon_{\text{Si}}(1-\varepsilon_{\text{Si}})}{(1+\varepsilon_{\text{Si}}) - f(1-\varepsilon_{\text{Si}})} \tag{1}$$

and
$$\varepsilon_{\perp,\text{eff}} = \varepsilon_{\text{Si}} + f(1-\varepsilon_{\text{Si}}) \tag{2}$$

Here, the subscripts "$\parallel$" and "$\perp$" respectively denote directions parallel and vertical to the SiNH-vacuum interface, and $f = \pi D^2/4P^2$ is the volumetric filling ratio. $\varepsilon_{\text{Si}}$ is the dielectric function of doped silicon which can be obtained by a Drude model [35]:

$$\varepsilon_{\text{Si}}(\omega, N, T) = \varepsilon_{\infty} - \frac{\omega_{\text{p}}^2}{\omega^2 + i\Gamma\omega} \tag{3}$$

where $\varepsilon_{\infty} = 11.7$ is the high-frequency constant, $\Gamma$ is the temperature-dependent scattering rate, and $\omega_{\text{p}} = \sqrt{N_c e^2 / m^* \varepsilon_0}$ is the plasma frequency with carrier concentration $N_c$, electron charge $e$, carrier effective mass $m^*$, and the permittivity of free space $\varepsilon_0$. Here, the effect of doping level is accounted by the carrier concentration which is the product of doping level and degree of ionization. As shown in Fig. 2(a), the D-SiNH emitter with $p$-type doping level $N_1 = 10^{20}$ cm$^{-3}$ and $f = 0.5$ exhibits uniaxial metallic behavior at frequencies $\omega < 2.3 \times 10^{14}$ rad/s where both $\varepsilon_{\perp,\text{eff}}$ and $\varepsilon_{\parallel,\text{eff}}$ are negative. Furthermore, the material property will change with different doping level $N_1$, and the region of uniaxial metallic behavior will shift towards higher frequency with increasing doping level (not shown here). Although the filling ratio $f$ can also tune material property of the D-SiNH emitter besides doping level $N_1$, it is fixed at $f = 0.5$ in the present study for simplicity.

## 2.2 Dielectric Response of Graphene

The dielectric function of graphene monolayer can be calculated by [29]

$$\varepsilon_{GR} = 1 + i\sigma / \omega\varepsilon_0 \Delta \tag{4}$$

where $\Delta = 0.5$ nm is the thickness of graphene, and the graphene conductivity $\sigma = \sigma_I + \sigma_D$ consists of interband and intraband (Drude) contributions [36, 37]:

$$\sigma_I(\omega,\mu) = \frac{e^2}{4\hbar}\left[G\left(\frac{\hbar\omega}{2}\right) + i\frac{4\hbar\omega}{\pi}\int_0^\infty \frac{G(\xi) - G(\hbar\omega/2)}{(\hbar\omega)^2 - 4\xi^2}d\xi\right] \tag{5a}$$

and

$$\sigma_D(\omega,\mu) = \frac{i}{\omega + i/\tau}\frac{2e^2 k_B T}{\pi\hbar^2}\ln\left[2\cosh\left(\frac{\mu}{2k_B T}\right)\right] \tag{5b}$$

where $\hbar$ is the reduced Planck constant, $k_B$ is the Boltzmann constant, $T$ is the absolute temperature, $\tau = 10^{-13}$ s is the relaxation time, $\mu$ is the chemical potential of graphene, and $G(\xi) = \sinh(\xi/k_B T)/[\cosh(\mu/k_B T) + \cosh(\xi/k_B T)]$. Figure 2(b) plots the real part of dielectric function of a free-standing graphene monolayer at different chemical potentials under a temperature of 300 K from $\omega = 1\times10^{13}$ to $3\times10^{14}$ rad/s. Clearly, the graphene with chemical potential $\mu > 0.1$ eV exhibits strong metallic behavior with Re($\varepsilon_{GR}$) < 0 within the entire spectral range of interests, while it shows dielectric behaviors with positive Re($\varepsilon_{GR}$) values at frequencies $\omega > 1.5\times10^{14}$ rad/s when $\mu = 0$ eV. When $\mu$ increases, the |Re($\varepsilon_{GR}$)| becomes larger, suggesting strong metallic behaviors dominated by graphene plasmon.

## 2.3. Near-Field Radiative Heat Transfer based on Fluctuational Electrodynamics Incorporated with Uniaxial Wave Propagation

The near-field radiative heat flux between two semi-infinite homogeneous media at temperatures of $T_1$ and $T_2$ ($T_1 > T_2$) can be calculated with fluctuational electrodynamics by [9]:

$$q = \frac{1}{\pi^2} \int_0^\infty d\omega \left[ \Theta(\omega, T_1) - \Theta(\omega, T_2) \right] \int_0^\infty s(\omega, \beta) d\beta \tag{6}$$

where $\beta$ is the parallel-component wavevector, and $\Theta(\omega, T) = \dfrac{\hbar\omega}{\exp(\hbar\omega/k_B T) - 1}$ is the mean energy of a Planck oscillator at the angular frequency $\omega$ and local thermal equilibrium temperature $T$. Since the contribution from propagating ($\beta < \omega/c$) and s-polarized waves are negligible when surface plasmon or phonon coupling exists [5, 9, 30], only the exchange function $s(\omega, \beta)$ for p-polarized evanescent ($\beta > \omega/c$) waves is considered here [38]:

$$s_{\text{evan}}(\omega, \beta) = \frac{\operatorname{Im}(r_{01}^p) \operatorname{Im}(r_{GR02}^p) \beta e^{-2\operatorname{Im}(\gamma_0)d}}{\left|1 - r_{01}^p r_{GR02}^p e^{i2\gamma_0 d}\right|^2} \tag{7}$$

where the subscripts 1, 0, 2 represent D-SiNH emitter, vacuum gap, and the graphene covered Si receiver, respectively. $\gamma_0 = \sqrt{\omega^2/c^2 - \beta^2}$ is the normal-component wavevector inside the vacuum. $r_{01}^p$ is the reflection coefficient for p-polarized waves at the vacuum-emitter interface with an expression of [6]:

$$r_{01}^p = \frac{\varepsilon_{1,\parallel}\gamma_0 - \gamma_1^p}{\varepsilon_{1,\parallel}\gamma_0 + \gamma_1^p} \tag{8}$$

where $\gamma_1^p = \sqrt{\varepsilon_{1,\parallel}\omega^2/c^2 - \varepsilon_{1,\parallel}\beta^2/\varepsilon_{1,\perp}}$ is the normal-component wavevector inside the D-SiNH emitter, and $\varepsilon_{1,\parallel}$ and $\varepsilon_{1,\perp}$ are respectively the in-plane and out-of-plane effective dielectric functions of D-SiNH described by Eqs. (1) and (2). Note that uniaxial wave propagation has to be considered here due to the anisotropic optical response of the D-SiNH emitter according to the effective medium approximation. On the other hand, $r^p_{GR02}$ is the modified reflection coefficient for p-polarized waves at the vacuum-receiver interface by treating the graphene as a surface current, which is given by [29]:

$$r^p_{GR02} = \frac{\omega\varepsilon_0(\varepsilon_2\gamma_0 - \gamma_2) + \sigma\gamma_0\gamma_2}{\omega\varepsilon_0(\varepsilon_2\gamma_0 + \gamma_2) + \sigma\gamma_0\gamma_2} \quad (9)$$

where $\varepsilon_2$ is the dielectric function of doped silicon plate as described by Eq. (3), and $\gamma_2 = \sqrt{\varepsilon_2\omega^2/c^2 - \beta^2}$ is the normal-component wavevector inside the receiver.

## 3. RESULTS AND DISCUSSION

### 3.1 Extraordinary Enhancement in Near-field Spectral Radiative Flux

In order to illustrate the near-field coupling effect between the D-SiNH emitter and the graphene-covered receiver, the spectral heat fluxes of three different cases are studied as shown in Fig. 3, where the Si plate receiver is lightly doped with $N_2 = 10^{15}$ cm$^{-3}$ and the vacuum gap $d$ = 20 nm for all three cases. Firstly, a heavily doped SiNH emitter with $N_1 = 10^{20}$ cm$^{-3}$ and a bare Si plate receiver are considered, which results in the spectral heat flux on the order of 10$^{-11}$ Wm$^{-2}$rad$^{-1}$s. Note that the maximum spectral radiative flux between two black bodies at the same temperatures of 400 K and 300 K is on the order of 10$^{-12}$ Wm$^{-2}$rad$^{-1}$s as shown in Fig. 3. However, by simply coating the Si plate receiver with a monolayer graphene at a chemical potential $\mu$ = 0.15 eV, the spectral heat flux is improved significantly with nearly two orders of magnitude enhancement. One major spectral peak with maximum $q_\omega$ = 6.8×10$^{-10}$ Wm$^{-2}$rad$^{-1}$s at the frequency of $\omega_1$ = 0.9×10$^{14}$ rad/s, while the other minor one with a peak value of $q_\omega$ = 9.4×10$^{-11}$ Wm$^{-2}$rad$^{-1}$s at $\omega_2$= 2.1×10$^{14}$ rad/s. The extraordinary spectral enhancement leads to an improvement of 72 times in the total heat flux over the blackbody limit. In order to examine whether the extraordinary enhancement in spectral heat flux is solely due to the graphene layer on the plate receiver or caused by the mutual coupling between the emitter and receiver, the doping level of the SiNH emitter is then reduced to $N_1 = 10^{15}$ cm$^{-3}$, while the receiver is still covered by the graphene sheet. It turns out that, the spectral heat flux between the lightly doped SiNH emitter and graphene-covered plate receiver is significantly reduced and becomes nearly the same as the first case with the heavily doped SiNH emitter and bare Si plate receiver without graphene. The two spectral peaks disappear as well. The comparison among all three cases clearly indicates that the extraordinary enhancement in near-field spectral heat flux is because of the coupling of the D-SiNH emitter and the graphene-covered receiver across the vacuum gap. As seen from Fig. 2, heavily doped SiNH emitter exhibits uniaxial metallic behaviors, while the graphene plasmon dominates its optical behaviors in the same frequency regime, which could lead to extraordinary photon tunneling and enhanced radiative transfer as a result of strong near-field SPP coupling across the vacuum gap. The following section will elucidate the underlying physical mechanism of the extraordinary SPP coupling between the D-SiNH and graphene-covered Si plate by means of exchange functions as well as analytical dispersion relations of coupled SPPs.

**3.2 Surface Plasmon Coupling between Dissimilar Materials**

Figure 4 shows the exchange function $s(\omega,\beta)$ between the heavily-doped SiNH emitter ($N_1 = 10^{20}$ cm$^{-3}$, $f = 0.5$) and the graphene-covered Si plate receiver ($N_2 = 10^{15}$ cm$^{-3}$, $\mu = 0.15$ eV) at the same vacuum gap of $d = 20$ nm. Note that the brighter contour indicates larger values of exchange function or stronger near-field photon tunneling. One bright and broad enhancement band (mode 1) can be clearly seen at the low-frequency regime, which is responsible for the major $q_\omega$ peak around $\omega_1 = 0.9\times10^{14}$ rad/s observed in Fig. 3. The excitation of the resonance mode 1 requires large normalized parallel wavevector $\beta^* = \beta c/\omega$ from 50 to 400. In the meantime, another relatively weaker enhancement (mode 2) in the $s(\omega,\beta)$ contour occurs around $\omega_2 = 2.1\times10^{14}$ rad/s at smaller $\beta^*$ values less than 200, which actually causes the minor spectral enhancement peak observed at higher frequencies. On the other hand, when the graphene is not present at the receiver surface or the emitter doping level becomes $N_1 = 10^{15}$ cm$^{-3}$, i.e., the first and third cases considered in Fig. 3, both enhancement modes in the exchange function disappear in either case (though not shown here), suggesting that both enhancements in the $s(\omega, \beta)$ are the results of strong coupling between the emitter and receiver across the nanometer vacuum gap.

In order to understand the underlying coupling mechanism between the D-SiNH emitter and the graphene-covered dielectric receiver, the dispersion relation of coupled SPPs across the vacuum gap can be analytically obtained by zeroing the denominator of the exchange function described in Eq. (9) as:

$$1 - r_{01}^p r_{GR02}^p e^{i2\gamma_0 d} = 0 \quad (10)$$

The coupled SPP dispersion curves are then solved and plotted along with the contour in Fig. 4. Excellent agreement is observed between the $s(\omega,\beta)$ contour enhancement and the coupled SPP dispersion, clearly verifying that the extraordinary photon tunneling between the D-SiNH and the graphene is due to the coupled SPPs, which is unusual between dissimilar materials.

Note that the optical properties of the D-SiNH would be significantly changed by the emitter doping level $N_1$ and filling ratio $f$, while those of the graphene-covered receiver would be largely altered by graphene chemical potential $\mu$ and the receiver doping level $N_2$. Therefore, the Fresnel reflection coefficients $r_{01}^p$ and $r_{GR02}^p$ at both vacuum interfaces would change, leading to the possible shifts of coupled SPP modes and thereby the near-field radiative transfer. The following two sections will study the effects of these parameters on the near-field photon tunneling in detail.

### 3.3 Effects of Graphene Chemical Potential $\mu$ and Emitter Doping Level $N_1$ on Near-Field Photon Tunneling

Let us first investigate the effects of graphene chemical potential $\mu$ and emitter doping level $N_1$ on the respective single-interface SPP without coupling across vacuum gap. $\mu$ affects the SPP resonance mode at the vacuum-receiver interface while $N_1$ impacts SPP at the emitter-vacuum interface. Note that the emitter filling ratio $f = 0.5$ and receiver doping level $N_2 = 10^{15}$ cm$^{-3}$ are kept unchanged. The SPP dispersion relation at the single graphene-covered vacuum-receiver interface can be calculated by zeroing the Fresnel reflection coefficient $r^p_{GR02}$:

$$\varepsilon_{2\parallel}\gamma_0^p + \gamma_2^p + \sigma\gamma_0^p\gamma_2^p/\varepsilon_0\omega = 0 \qquad (11)$$

Figure 5 shows how the single receiver interface SPP dispersion changes as a function of frequency $\omega$ and normalized parallel wavevector $\beta^* = \beta c/\omega$ when $\mu$ varies from 0 to 0.5 eV. Clearly, a larger graphene chemical potential will result in the SPP peak, where the two SPP modes merge, to shift towards higher frequencies. Note that higher $\beta^*$ values indicate more channels of photon tunneling or radiative transfer modes. The maximum $\beta^*_{max}$ is 78 at $\omega = 0.7\times10^{14}$ rad/s with $\mu = 0$ eV, and further increases to 139 at $\omega = 2.28\times10^{14}$ rad/s with $\mu = 0.1$ eV. Within the frequency range of interests, the number of radiative transfer modes becomes less with further increase of $\mu$.

On the other hand, the SPP dispersion at the single emitter-vacuum interface can be obtained by solving $r^p_{01} = 0$ or equivalently

$$\varepsilon_{1\parallel}\gamma_0 + \gamma_1^p = 0 \tag{12}$$

As shown in Fig. 5(b), the single SPP at the SiNH interface shows strong dependence on the emitter doping level $N_1$. Larger $N_1$ values would result in the SPP mode to shift towards higher frequencies. For a given doping level, the SPP dispersion shows little selectivity on angular frequencies with small $\beta^*$ values less than 2, suggesting few radiative transfer modes. The number of near-field photon tunneling channels significantly increases as seen via the abrupt increase of $\beta^*$ values when it approaches the asymptotic frequency, e.g., $\omega_{asy} = 2.1 \times 10^{14}$ rad/s for $N_1 = 10^{20}$ cm$^{-3}$. In general, the asymptotic frequency can be obtained under the condition $\text{Re}(\varepsilon_\perp \varepsilon_\parallel) = 1$, which can be analytically derived from the Eq. (12) when $\beta^* \gg 1$ for a uniaxial medium.

Now let us study the effects of $\mu$ and $N_1$ on the near-field SPP coupling. Figures 6(a) and 6(b) present the exchange functions between the D-SiNH emitter ($N_1 = 10^{20}$ cm$^{-3}$) and graphene-covered receiver across a 20-nm vacuum gap at different graphene chemical potentials of $\mu = 0.3$ and 0.5 eV, respectively. In comparison with the exchange function at $\mu = 0.15$ eV shown in Fig. 4, it can be clearly seen that, larger $\mu$ values cause the shift of the enhancement mode 1 in $s(\omega,\beta)$ towards higher frequencies. Although $\beta^*$ values becomes smaller, the strength of the exchange function actually increases with larger $\mu$. The $\mu$ effect on the coupled SPP shifting is consistent with that on the single-interface SPP shown in Fig. 5(a). On the other hand, the enhancement mode 2 barely changes by the $\mu$. The behaviors of both SPP coupling modes 1 and 2 at different $\mu$ are further verified by the excellent agreement between the coupled SPP dispersion and the enhancement in exchange function. The SPP dispersion curves enhancement mode 2 also indicates an asymptotic frequency of $2.1\times10^{14}$ rad/s, which matches well with that of the SPP at the single emitter-vacuum interface with $N_1 = 10^{20}$ cm$^{-3}$ as indicated in Fig. 5(b).

On the other hand, when reducing the $N_1$ values to $10^{19}$ cm$^{-3}$ and $10^{18}$ cm$^{-3}$ as shown respectively in Figs. 6(c) and 6(d), the enhancement in exchange function changes significantly. As learned from the single-interface SPP behavior, the asymptotic frequency of the SPP at the emitter interface would shift toward lower frequencies when $N_1$ decreases, resulting in the coupled SPPs associated with the enhancement mode 2 to be pushed towards lower frequencies and couple with the enhancement mode 1 region. As a result, with $N_1 = 10^{19}$ cm$^{-3}$ the four coupled modes which occurs within the same frequency region strongly interact with each other, leading to a merged and much stronger enhancement in the exchange function contour. When $N_1$ further decreases to $10^{18}$ cm$^{-3}$, the enhancement mode 2 shifts to even lower frequencies than the enhancement mode 1, as seen by the coupled SPP dispersion curves, resulting in a much broader enhancement in $s(\omega,\beta)$. Note that, the enhancement mode 1 region due to the coupled SPP little shifts with $N_1$. Clearly, the emitter doping level $N_1$ or the surface plasmon at the emitter-vacuum interface dominates the enhancement mode 2 on $s(\omega,\beta)$ contour, while graphene chemical potential $\mu$ or the surface plasmon at the vacuum-graphene interface plays a significant role only in the enhancement mode 1 due to SPP coupling. By manipulating the $N_1$ and $\mu$ values, all the coupled SPP modes could interact with each other at different strengths, which could significantly change the exchange function and thereby modulate the near-field radiative transfer.

In order to quantitatively demonstrate the near-field coupling effect on the photon tunneling, a near-field enhancement factor is defined as $\Xi = q/q_{BB}$, where $q$ is the total near-field radiative flux between the D-SiNH emitter and graphene-covered planar receiver integrated over frequency range from $1\times10^{13}$ rad/s to $3\times10^{14}$ rad/s, and $q_{BB} = 992$ W/m$^2$ is the radiative flux between two black bodies at temperatures of 400 K and 300 K respectively. Figure 7 shows the near-field enhancement factor as a function of chemical potential $\mu$ from 0 to 1 eV at different emitter doping levels $N_1$ from $10^{15}$ to $10^{21}$ cm$^{-3}$. The receiver doping level $N_2$ is $10^{15}$ cm$^{-3}$ and the vacuum gap is $d = 20$ nm. Firstly of all, when $N_1 = 10^{20}$ cm$^{-3}$, the enhancement factor is only about 55 at $\mu = 0$ eV, starts to increase at $\mu = 0.1$ eV, reaches about 370 at $\mu = 0.6$ eV, and then saturates with larger $\mu$ values. The $\mu$ effect on the near-field radiative flux enhancement can be clearly understood by the exchange functions in Figs. 6(a) and 6(b), where the two enhancement regions of different SPP modes couple with each other. As a result of stronger coupling, the exchange function and thereby the radiative flux become larger. Note that, after the two modes are coupled together, further increment of chemical potential will only result in saturation of heat flux. The $\mu$ effect on the near-field radiative flux is similar with more doping at $N_1 = 10^{21}$ cm$^{-3}$, while only 40 times enhancement is achieved over the blackbody limit. This is because the asymptotic frequency of the SPP at the emitter interface is at $7\times10^{14}$ rad/s shown in Fig. 5(b), which is much higher than the resonance frequencies of the major coupled SPP modes (i.e., mode 1) dominated by the receiver interface. Therefore, the coupling strengths of the two modes at large $\mu$ is much weaker compared to that with $N_1 = 10^{20}$ cm$^{-3}$.

Interestingly, the $\mu$ effect on the enhancement factor is different for $N_1 \leq 10^{19}$ cm$^{-3}$. There occurs a maximum enhancement at $N_1 = 10^{19}$ cm$^{-3}$ and $\mu = 0.15$ eV, which is about 460 times higher than the heat flux between two blackbodies, while the enhancement becomes smaller at either smaller or larger $\mu$. This is actually not surprising after the exchange function in Fig. 6(c) is understood, where strongest coupling between the two modes occurs. Different $\mu$ values will shift the enhancement mode 1 which is dominated by the receiver interface to different frequencies that are away from the asymptotic frequency of the SPP at the emitter interface, leading to weaker SPP coupling and thus smaller near-field radiative heat flux. On the other hand, when $N_1$ is further decreased, the asymptotic frequency would shift towards lower frequencies less than $1\times10^{13}$ rad/s indicated by Fig. 5(b). As a consequence, the surface plasmon at the emitter interface cannot couple efficiently with the other graphene-dominated interface, whose SPP occurs at higher frequencies. Thus, it is understandable that the enhancement factor for $N_1 = 10^{18}$ and $10^{17}$ cm$^{-3}$ is high only at $\mu = 0$ eV, and then monotonically decreases with larger $\mu$ values, which actually shifts the enhancement mode 1 towards higher frequencies further away from the enhancement mode 2 at low frequencies. For lightly doped SiNH emitter with $N_1 \leq 10^{16}$ cm$^{-3}$, the surface plasmon at two interfaces cannot couple at all across the nanometer vacuum gap, resulting in the lowest radiative flux, and the graphene chemical potential could barely take into any effect.

**3.4 Effect of Receiver Doping Level $N_2$ on Near-Field Photon Tunneling**

It should be noted that, besides $N_1$ and $\mu$, the receiver doping level $N_2$, which changes the reflection coefficient $r_{GR02}^p$ at the vacuum-graphene-D-Si plate interface, could potentially affect the near-field photon tunneling as well. Figure 8 presents the near-field enhancement factor as a function of $\mu$ at different $N_2$ values with the emitter doping level $N_1$ is fixed at $10^{20}$ cm$^{-3}$ and $d = 20$ nm. After low doping levels from $10^{15}$ to $10^{18}$ cm$^{-3}$, $N_2$ does have little effect on the normalized radiative heat flux. On the other hand, the enhancement factor $\Xi$ first does not change with $\mu$, then starts to monotonically increase at $\mu = 0.1$ eV, and finally saturates at 400 for $\mu > 0.5$ eV. Basically at low doping level $N_2$, which has almost no effect on tuning the surface plasmon at the graphene-covered interface, graphene chemical potential dominates and the near-field coupling between the two modes becomes stronger with larger $\mu$ as explained previously. $N_2$

starts to play a role in further enhancing the radiative flux up to about 500 times over the blackbody limit when it increases to $10^{19}$ cm$^{-3}$, at which the $\mu$ effect is similar with that at less doping in the silicon plate.

However, when $N_2$ further increases to $10^{20}$ cm$^{-3}$, enhancement factors around 500 is achieved with small $\mu$ values less than 0.1 eV, and starts to monotonically decrease when $\mu$ becomes larger. The different trend of the $\mu$ effect can be understood by the effect of $N_2$ on tuning the surface plasmon of doped silicon plates. As studied in Ref. [39], higher doping level increases the charge density of doped silicon and pushes the plasmon frequency towards higher frequencies into the infrared region under investigation. In other words, larger doping level $N_2$ could shift the coupled SPP mode 1 toward higher frequencies but have no effect on the emitter-dominated mode 2, as demonstrated by the dispersion curves in Fig. 9(a) in comparison with Fig. 4. Therefore, even at small $\mu$ like 0.1 eV in this case, both coupled SPP modes could already occurs at nearby frequencies, and strong coupling and merging could occur. When further increasing the $\mu$, the coupled enhancement mode 1 shifts to higher frequencies, resulting in decreased coupling strength between the two modes, as demonstrated by Fig. 9(b) with $\mu = 0.5$ eV, and thereby smaller radiative flux. If the silicon plate receiver is heavily doped with $N_2 = 10^{21}$ cm$^{-3}$, the coupled mode 1 will occur at higher frequencies than the coupled mode 2 even at $\mu = 0$, in which case the coupling between the two modes are already weak. Larger $\mu$ values, which pushes the mode 1 towards even higher frequencies, will not strengthen but keep weakening the coupling of two modes, leading to much smaller near-field enhancement factor around 50 times with little $\mu$ effect, as can clearly be seen in Fig. 8.

**3.5 Dependence of Vacuum Gap Distance on Near-field Photon Tunneling**

Finally, the effect of vacuum gap on the near-field photon tunneling is investigated. Figure 10 shows the near-field enhancement factor $\Xi$ at different vacuum gap distance $d$ from 10 nm up to 1 μm with varying graphene chemical potential $\mu$ from 0 eV to 1 eV. The doping levels for the SiNH emitter and the silicon plate receiver are $N_1 = 10^{20}$ cm$^{-3}$ and $N_2 = 10^{15}$ cm$^{-3}$, respectively. Overall, the near-field radiative transfer decreases with larger vacuum gap, which is understandable as the coupled SPP modes becomes weaker when the emitter and receiver further apart. When $d > 500$ nm the near-field enhancement factor approaches one, in other words, the radiative transfer goes to the far-field limit. As a matter of fact, the graphene dominated resonant mode 1 disappears when gap distance is larger than 200 nm, while the SiNH dominated resonant mode 2 vanishes when $d > 70$ nm, both of which are confirmed by exchange function plots (not shown here). The near-field radiative transfer is greatly enhanced at $d < 70$ nm, where both resonant modes occur and could further couple with each other. More than two orders of magnitude enhancement over the blackbody limit can be achieved at vacuum gaps less than 40 nm, in which case larger graphene chemical potentials greatly promotes the extraordinary near-field photon transport. Note that up to three orders of magnitude enhancement could be achieved between the considered dissimilar materials at sub-20-nm vacuum gaps with proper tuning of graphene chemical potential between 0.2 and 0.8 eV. However, cautions need to be paid here as the effective medium approximation for the SiNH emitter may no longer be valid depending on the comparison of the pattern size such as period $P$ to the vacuum gap $d$, as EMT is generally valid for near-field radiative transfer only when $P \ll \pi d$. Here, the discussion is limited to the smallest vacuum gap of 10 nm, because nonlocal effect for the dielectric response of materials would occur at sub-10-nm regime, which the present study did not consider.

## 4. CONCLUSION

In summary, near-field radiative flux between the doped SiNH and graphene-covered silicon plate is theoretically studied. The extraordinary photon transport is analyzed in detail as due to unusual coupled surface plasmon coupling between the uniaxial metallic metamaterial emitter and graphene-covered receiver. With the help of fluctuational electrodynamics, it is shown that two coupled surface plasmon modes, which are respectively dominated by the SiNH-vacuum and the vacuum-graphene-silicon interfaces, will shift and interact with each other under different doping levels ($N_1$ and $N_2$), and graphene chemical potential $\mu$. The different coupling strength between the surface plasmon modes could lead to either enhancement or suppression of near-field radiative heat transfer between the dissimilar materials considered here. With proper tuning of $N_1$, $N_2$, and $\mu$ values, the near-field heat flux reaches up to 500 times higher than that between two black bodies at vacuum gap $d = 20$ nm. The results and understanding gained here will facilitate the exploration and application of novel metamaterials for energy conversion and thermal management by means of near-field photon transport.

**Acknowledgements**

Supports from the National Science Foundation under CBET-1454698 (YY and LW) and ASU New Faculty Startup fund (JYC) are greatly acknowledged.

**Figure Captions:**

Fig. 1   Schematic of the simulated structure separated by vacuum gap $d$ where both the doped SiNH emitter and graphene covered D-Si receiver are assumed to be semi-infinite.

Fig. 2   The real part of (a) vertical and parallel SiNH dielectric functions ($N_1 = 10^{20}$ cm$^{-3}$) and (b) graphene dielectric functions with respect to different chemical potential $\mu$.

Fig. 3   The spectral heat flux of four different setups: SiNH emitter ($N_1 = 10^{20}$ cm$^{-3}$) with D-Si receiver ($N_2 = 10^{15}$ cm$^{-3}$), SiNH emitter ($N_1 = 10^{20}$ cm$^{-3}$) with graphene covered D-Si receiver ($\mu = 0.15$ eV, $N_2 = 10^{15}$ cm$^{-3}$), SiNH emitter ($N_1 = 10^{15}$ cm$^{-3}$) with graphene covered D-Si receiver ($\mu = 0.15$ eV, $N_2 = 10^{15}$ cm$^{-3}$), and two blackbodies as a function of angular frequency.

Fig. 4   The exchange function between SiNH emitter ($N_1 = 10^{20}$ cm$^{-3}$) and graphene ($\mu = 0.15$ eV) covered D-Si receiver ($N_2 = 10^{15}$ cm$^{-3}$) separated by a vacuum gap of $d = 20$ nm. The blue and yellow dashed curves (mode 1 and 2) represent the coupled SPP dispersion curves dominated by vacuum-graphene and SiNH-vacuum interfaces, respectively.

Fig. 5   The single interface SPP between (a) vacuum and graphene covered D-Si receiver ($N_2 = 10^{15}$ cm$^{-3}$) with respect to different graphene chemical potential and (b) vacuum and SiNH of different doping level $N_1$.

Fig. 6   The exchange function between SiNH emitter ($N_1 = 10^{20}$ cm$^{-3}$) and graphene covered D-Si receiver ($N_2 = 10^{15}$ cm$^{-3}$) with graphene chemical potential of (a) $\mu = 0.3$ eV and (b) $\mu = 0.5$ eV; and that between SiNH emitter and graphene ($\mu = 0.15$ eV) covered D-Si receiver ($N_2 = 10^{15}$ cm$^{-3}$) with emitter doping level of (c) $10^{19}$ cm$^{-3}$ and (d) $10^{18}$ cm$^{-3}$. The vacuum gap $d$ is fixed at 20 nm.

Fig. 7   The NFR enhancement factor $\Xi$ as a function of graphene chemical potential with respect to different emitter doping level $N_1$ while the receiver has a doping level of $N_2 = 10^{15}$ cm$^{-3}$ and the gap distance is set to be $d = 20$ nm.

Fig. 8    The NFR enhancement factor $\Xi$ as a function of graphene chemical potential with respect to different receiver doping level $N_2$ while the emitter has a doping level of $N_1 = 10^{20}$ cm$^{-3}$ and the gap distance is set to be $d = 20$ nm.

Fig. 9    The exchange function between SiNH emitter ($N_1 = 10^{20}$ cm$^{-3}$) and graphene covered D-Si receiver ($N_2 = 10^{20}$ cm$^{-3}$) with graphene chemical potential of (a) 0.1 eV and (b) 0.5 eV while the gap distance is set to be $d = 20$ nm.

Fig. 10   The impact of vacuum gap distance $d$ and graphene chemical potential $\mu$ on near-field enhancement factor $\Xi$ under the emitter doping level $N_1 = 10^{20}$ cm$^{-3}$ and receiver doping level $N_2 = 10^{15}$ cm$^{-3}$.

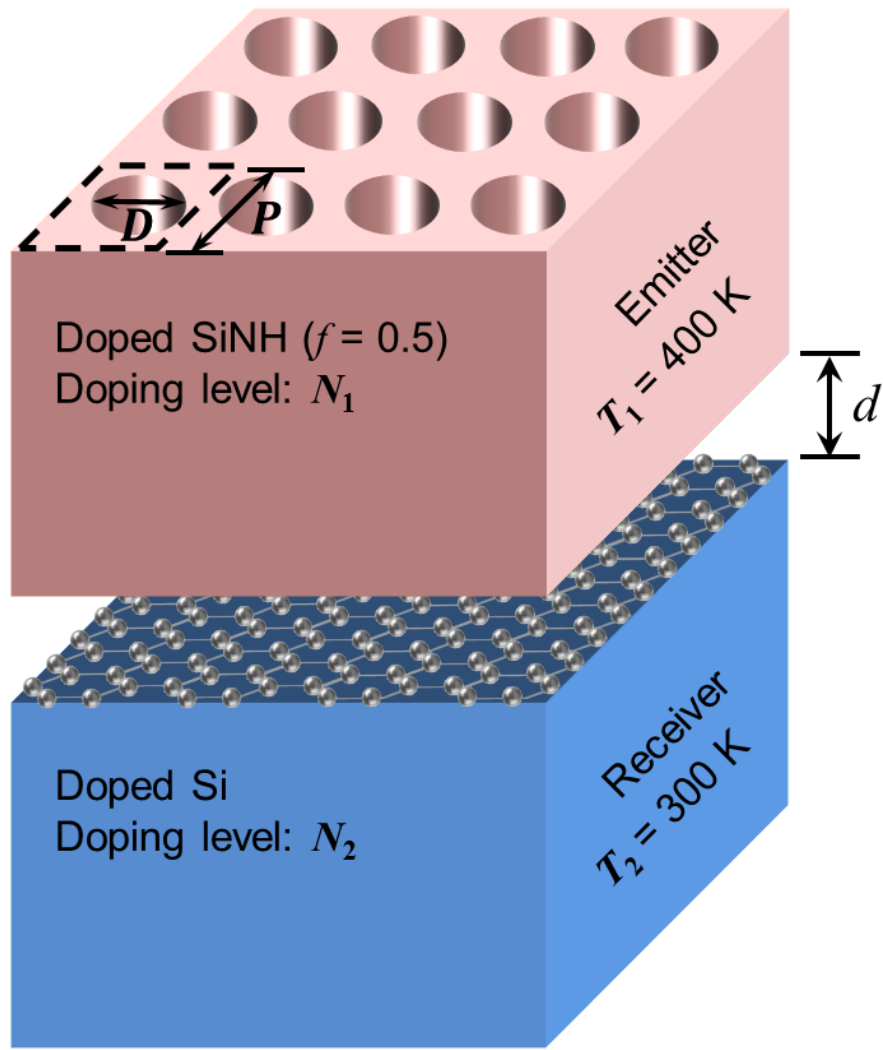

**Chang et al, Fig. 1**

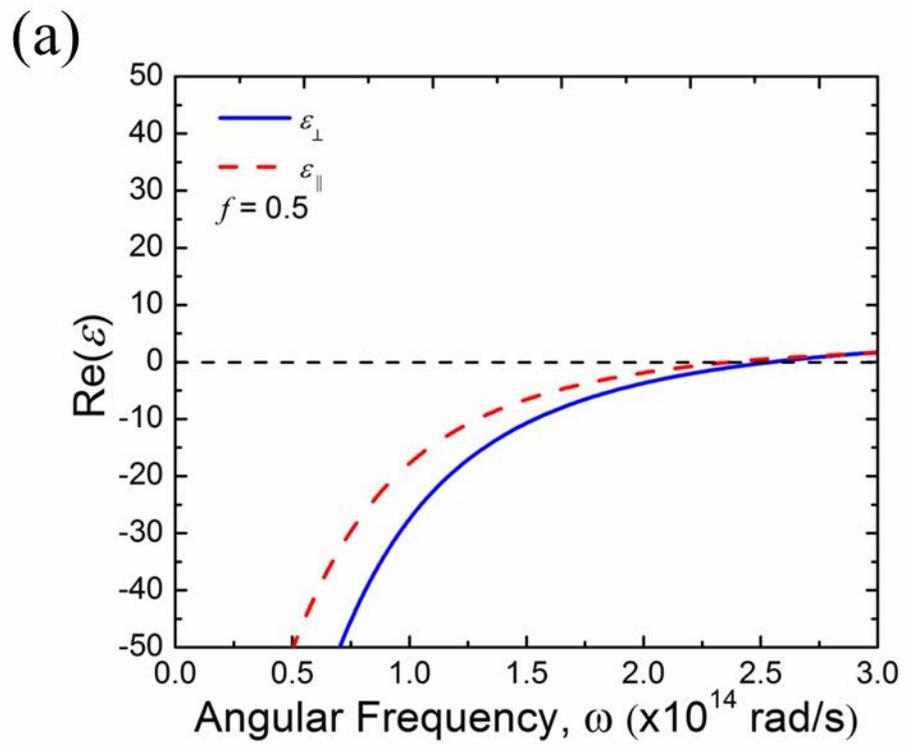

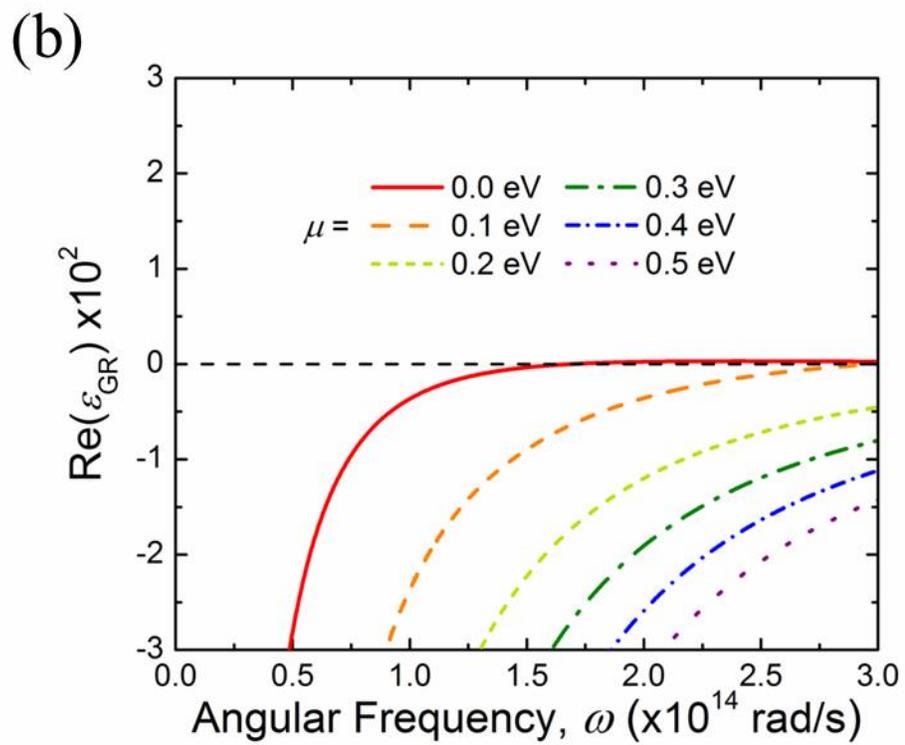

**Chang et al, Fig. 2**

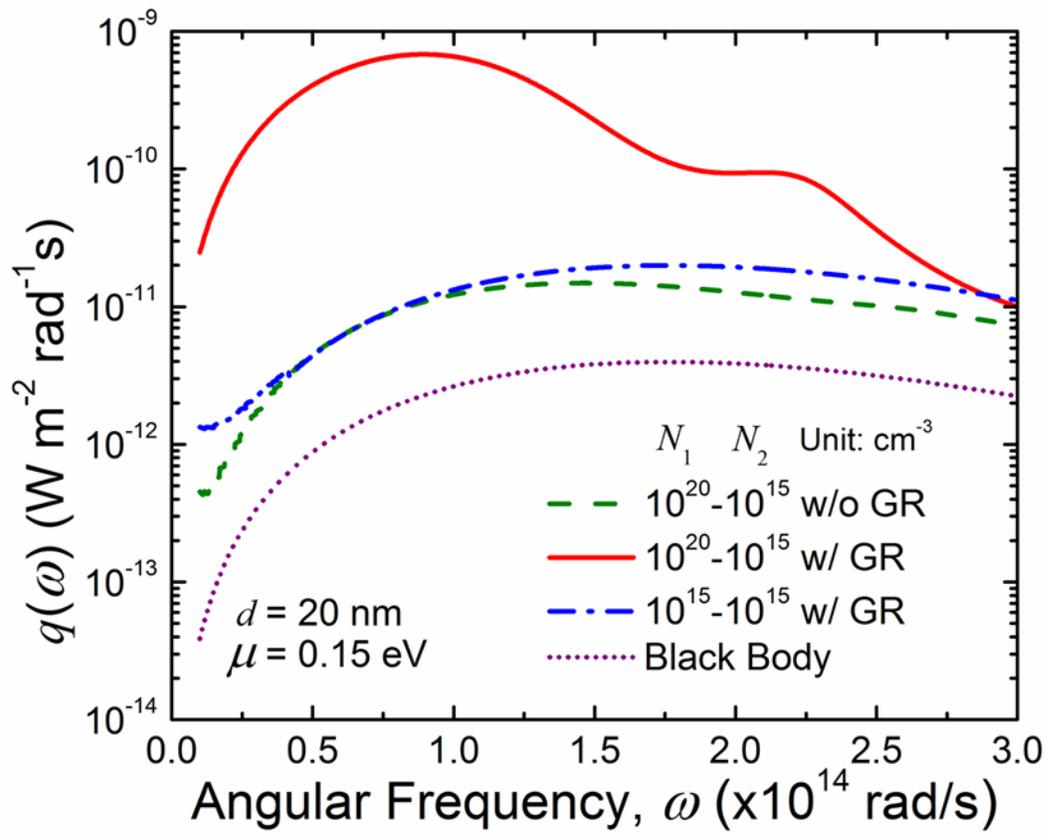

**Chang et al, Fig. 3**

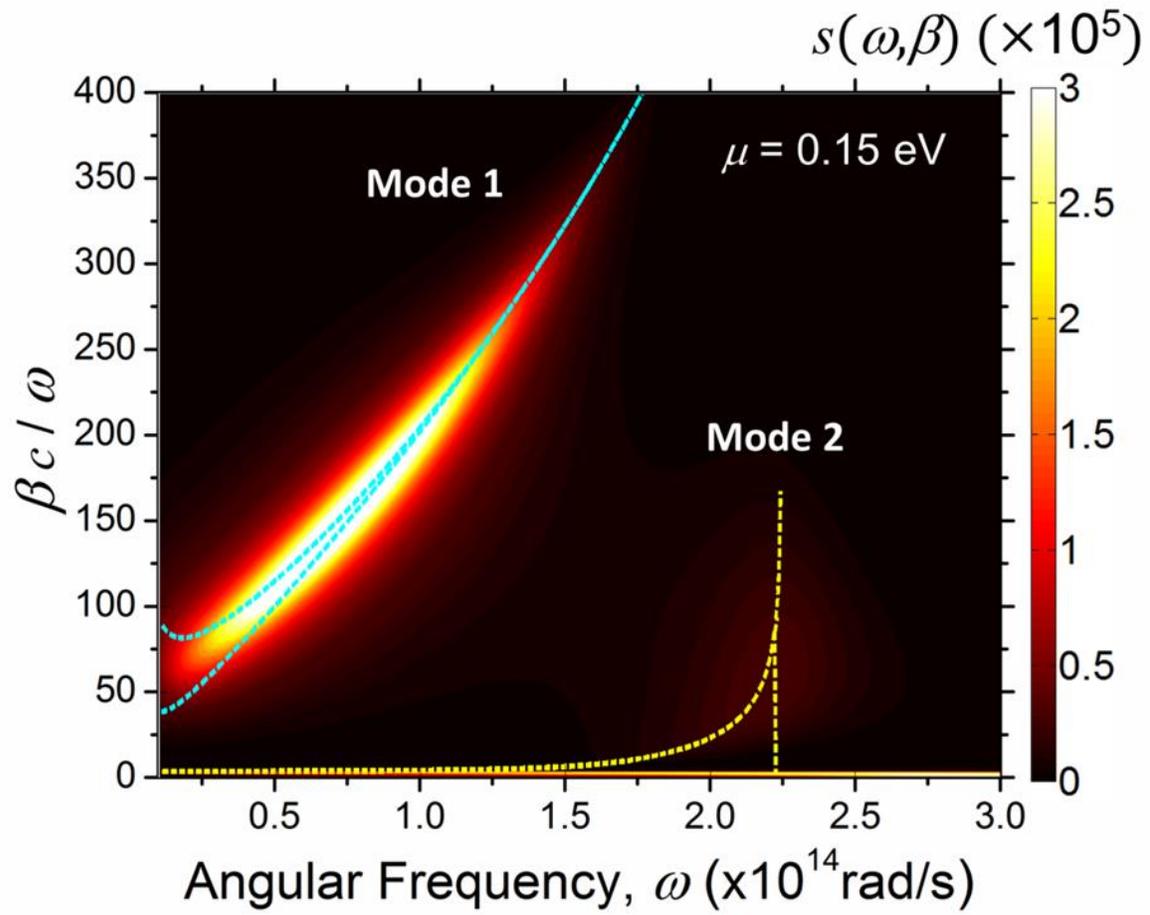

**Chang et al, Fig. 4**

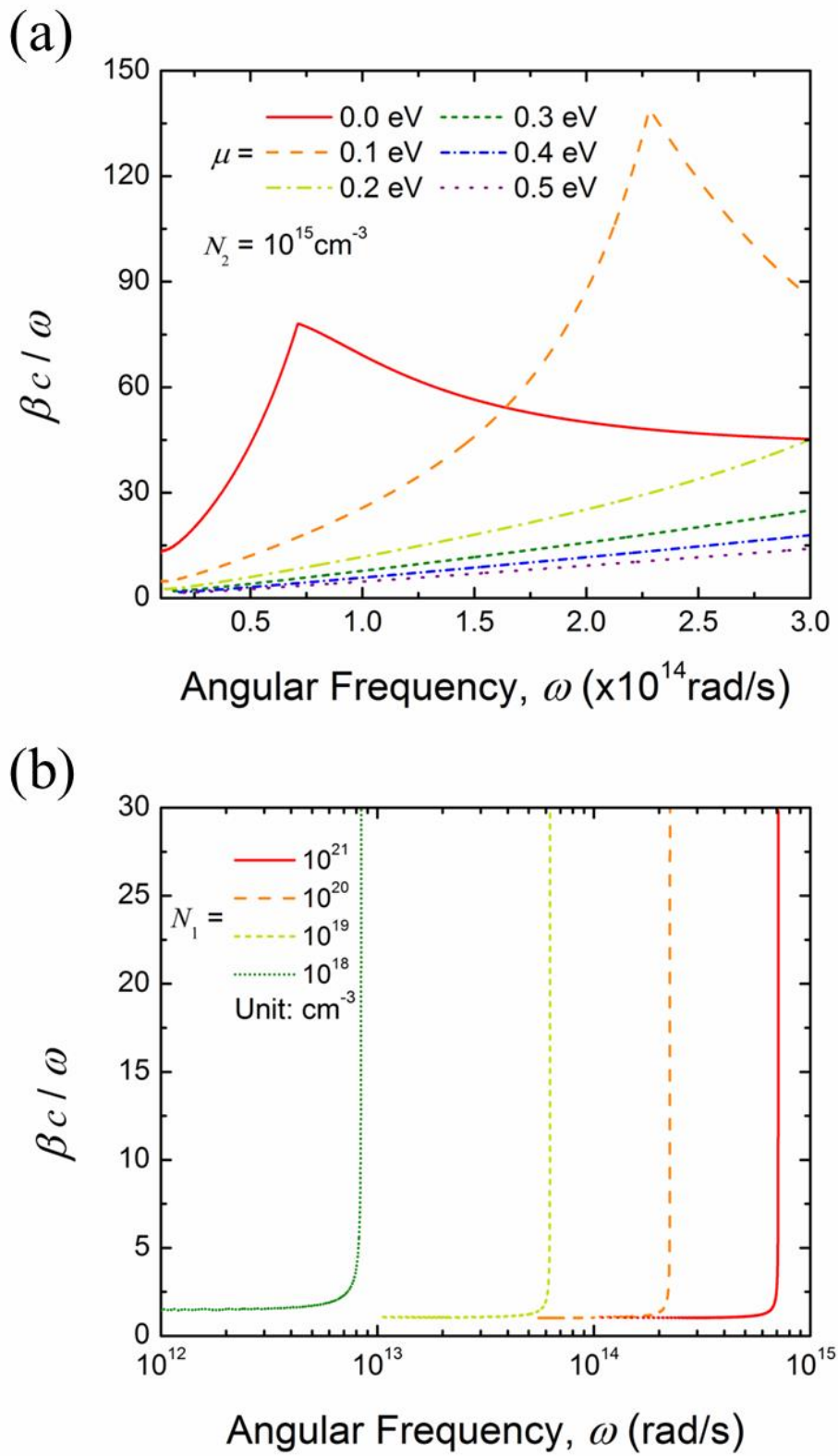

**Chang et al, Fig. 5**

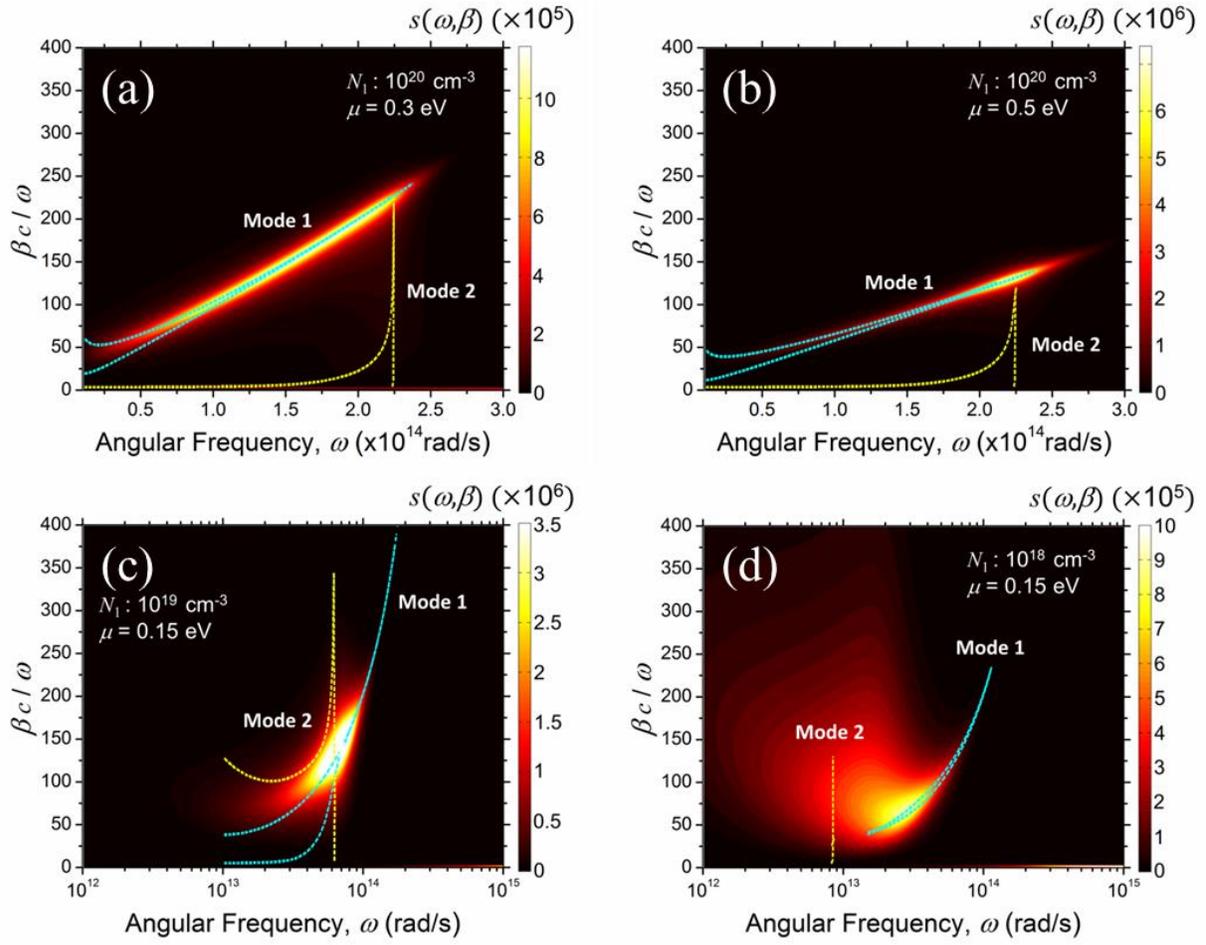

**Chang et al, Fig. 6**

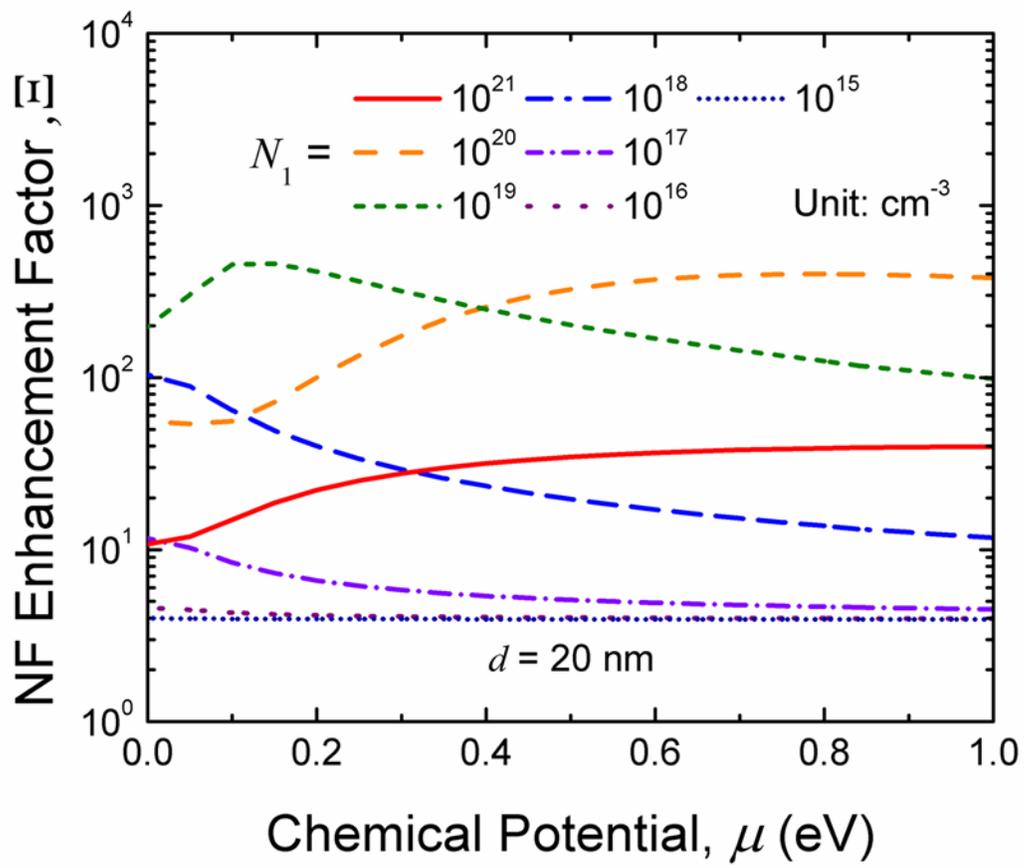

**Chang et al, Fig. 7**

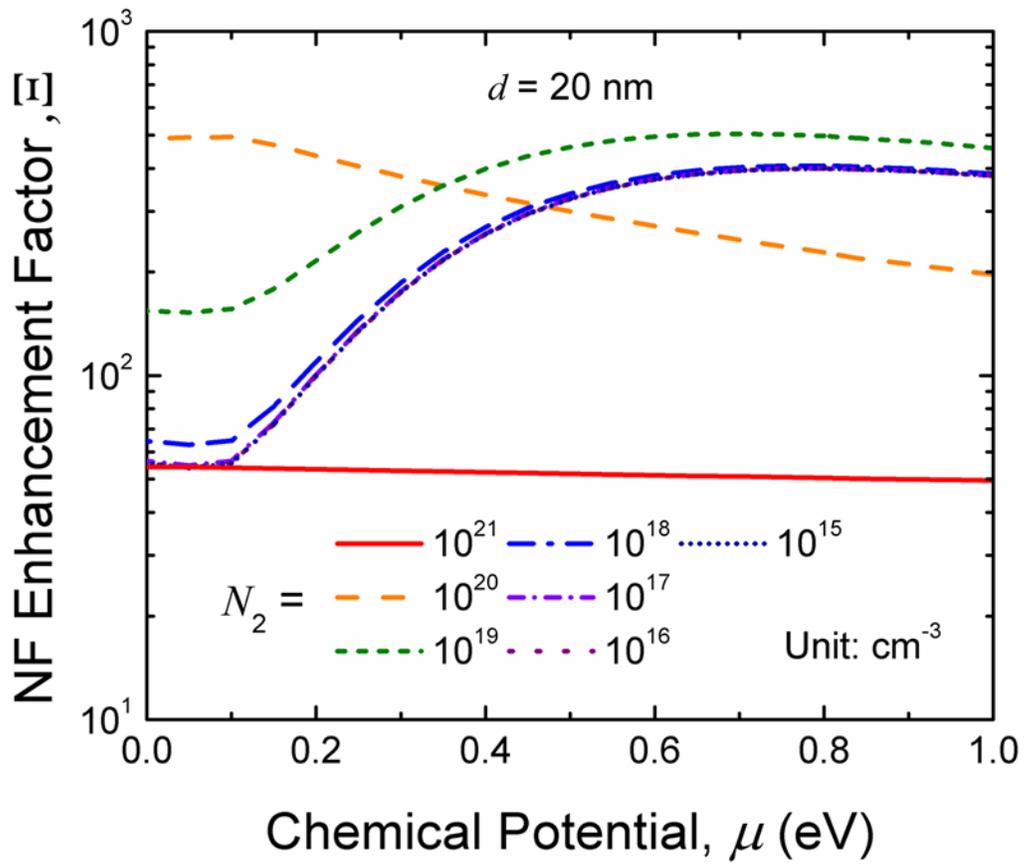

**Chang et al, Fig. 8**

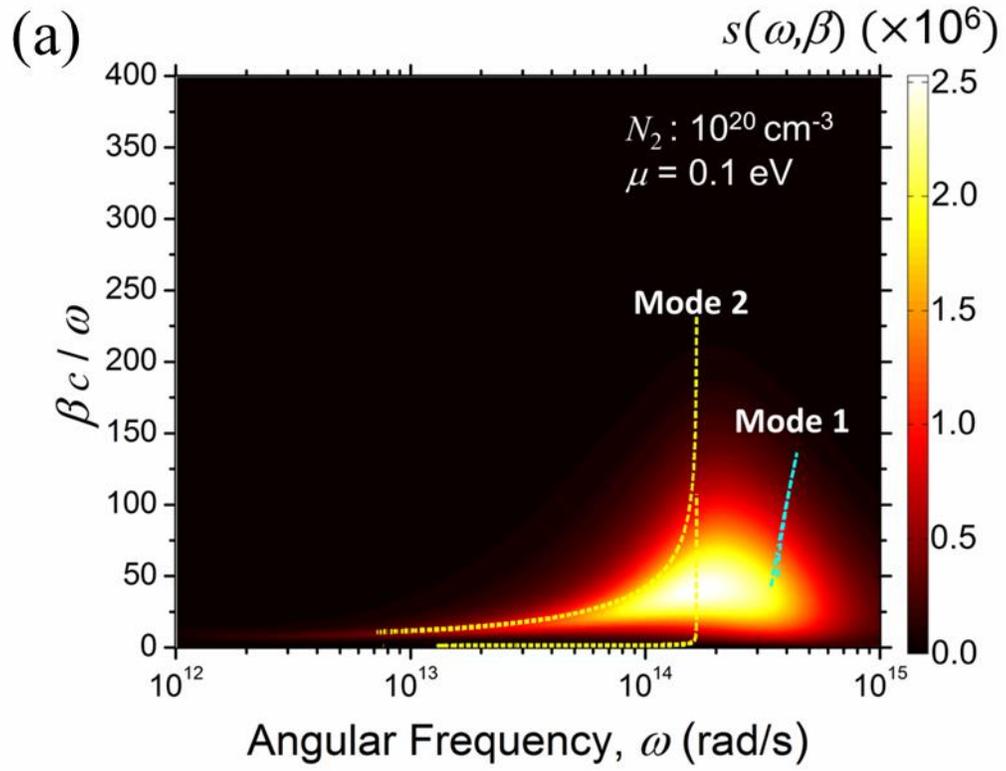

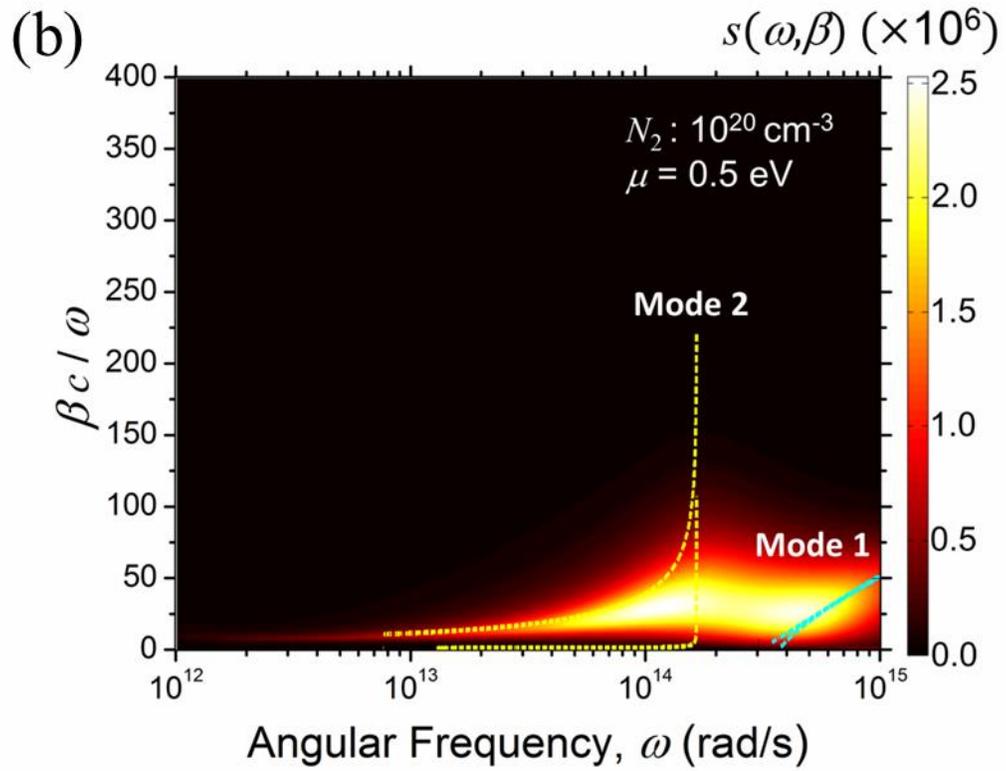

**Chang et al, Fig. 9**

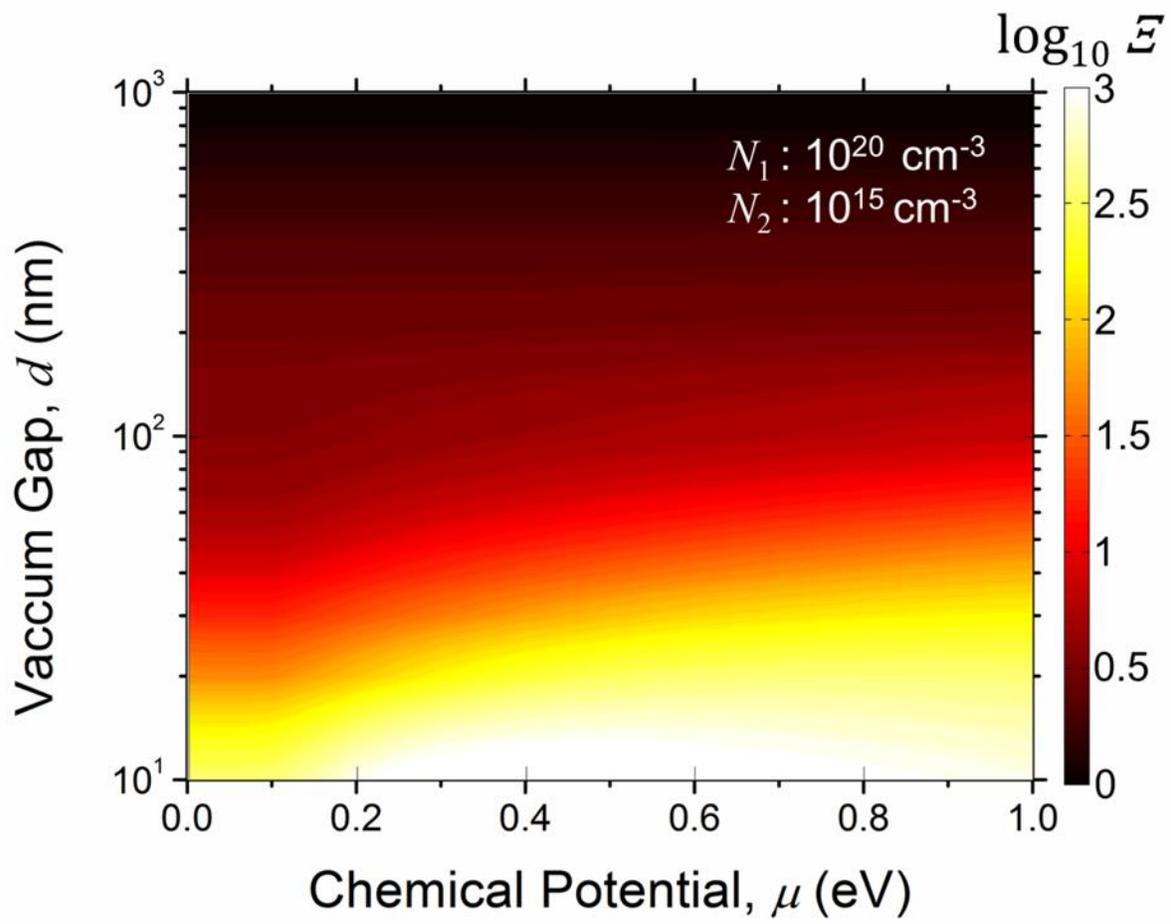

**Chang et al, Fig. 10**